\documentclass[sigconf]{acmart}
\AtBeginDocument{%
  \providecommand\BibTeX{{%
    \normalfont B\kern-0.5em{\scshape i\kern-0.25em b}\kern-0.8em\TeX}}}

\setcopyright{acmcopyright}
\copyrightyear{2023}
\acmYear{2023}
\acmDOI{XXXXXXX.XXXXXXX}

\acmConference[CHI'23]{Make sure to enter the correct
  conference title from your rights confirmation email}{April 23--28,
  2023}{Hamburg, Germany}
\acmBooktitle{Hamburg '23: ACM XXX,
 April 23--28, 2023, Hamburg, Germany} 
\acmPrice{0.00}
\acmISBN{978-1-4503-XXXX-X/18/06}

\newcommand{\toolname}{DataGarden}

\begin{document}

\title[DataGarden]{DataGarden: Exploring Our Community in a VR Data Visualization}

\author{Joy Kondo}
\authornote{Both authors contributed equally to this research.}
\affiliation{%
  \institution{Boston College}
  \city{Chestnut Hill}
  \country{USA}}
 \email{joy.kondo@bc.edu}
 
\author{Justin Park}
\authornotemark[1]
\affiliation{%
  \institution{Boston College}
  \city{Chestnut Hill}
  \country{USA}}
 \email{justin.park@bc.edu}
 
\author{Josiah Kondo}
\affiliation{%
  \institution{Boston College}
  \city{Chestnut Hill}
  \country{USA}}
\email{josiah.kondo@bc.edu}
 
\author{Nam Wook Kim}
\affiliation{%
  \institution{Boston College}
  \city{Chestnut Hill}
  \country{USA}}
\email{nam.wook.kim@bc.edu}

\renewcommand{\shortauthors}{Kondo and Park, et al.}

\begin{abstract}
As our society is becoming increasingly data-dependent, more and more people rely on charts and graphs to understand and communicate complex data. While such visualizations effectively reveal meaningful trends, they unavoidably aggregate data into points and bars that are overly simplified depictions of ourselves and our communities. We present DataGarden, a system that supports embodied interactions with humane data representations in an immersive VR environment. Through the system, we explore ways to rethink the traditional visualization approach and allow people to empathize more deeply with the people behind the data. 

\end{abstract}



\begin{teaserfigure}
  \centering  \includegraphics[width=\textwidth]{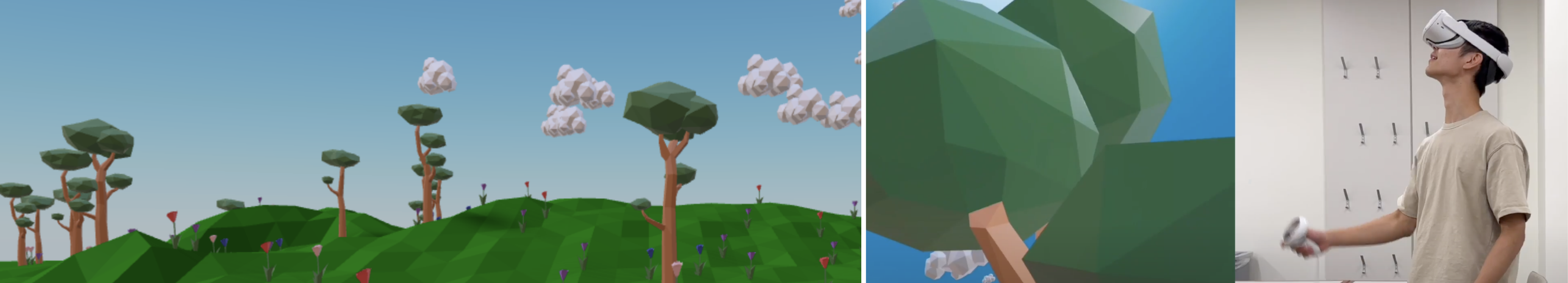}
  \caption{An overview of the \toolname{} virtual environment. Each flower or tree represents an individual person, and the various characteristics including the flower color and number of clouds are determined by users' survey responses.}
  \Description{Wide view showing the DataGarden virtual environment. A focused view of a tree. A person using a Meta Quest device to look up the tree.}
  \label{fig:teaser}
\end{teaserfigure}


\maketitle

\section{Introduction}
Traditional approaches to data visualization have commonly relied on abstract visualizations, such as bars, lines, and pies, to objectively represent data in a minimalist manner. While these methods have demonstrated perceptual effectiveness in conveying statistical information, such abstract representations tend to overlook the human aspect underlying the data~\cite{boy2017showing}. Moreover, conventional design guidelines such as avoidance of visual clutter have reduced the stories and emotions represented by data visualizations to mere graphical marks \cite{hullman2011benefitting}. However, given that the data we gather from individuals' lived experiences offers a glimpse into our world and the people it comprises, it is crucial that data visualizations elicit an emotional connection with the individuals represented by the data~\cite{giorgia2016data}.

In this paper, we present \toolname{}, a novel data visualization system leveraging immersiveness and humanistic representations of data to help people reflect on the people behind the data. Our system is based on the empirical foundations demonstrating the benefits of immersion and pictorial representations~\cite{borkin2015beyond,romat2020dear}. Designers can collect qualitative questionnaire data and construct mappings from the data to virtual objects and their visual properties in \toolname{}. Users can then use VR devices like Meta Quest to interact with the data in an immersive environment. The current system supports garden objects such as trees and flowers, but the visual vocabulary can be expanded and customized to incorporate additional visual metaphors. The closest work to ours is DataSelfie~\cite{kim2019dataselfie}, but we extend the idea to a VR environment and community data rather than 2D and individual responses. Below, we describe our design goals and system components in detail. 

\begin{figure*}[t]
  \centering
  \includegraphics[width=\linewidth, height=3cm]{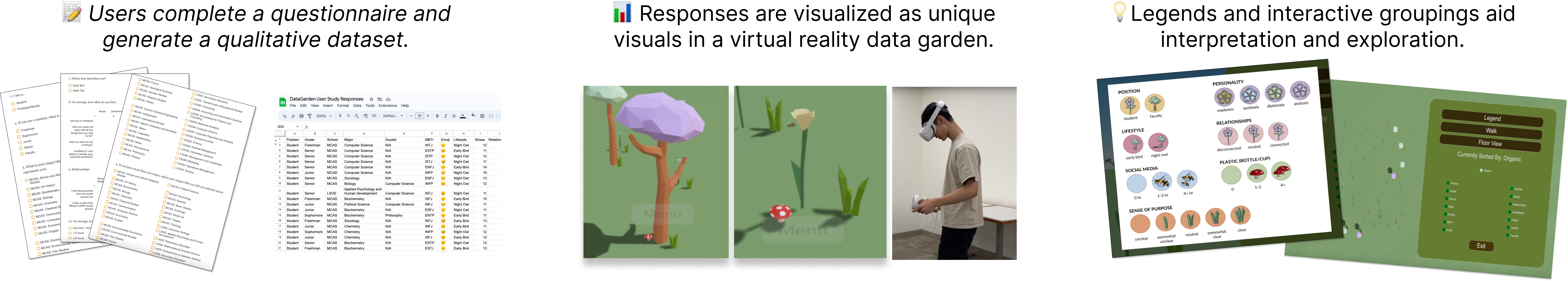}
  \caption{DataGarden's data visualization pipeline. Left: personal and community data is collected via a questionnaire. Middle: responses are mapped to unique visual objects. Right: legends and interactive groupings facilitate data interpretation. }
  \Description{}
\end{figure*}

\section{DataGarden Design and System}

\subsection{Design Goals}

\subsubsection{Capturing human elements in data}: we want to collect data that contains rich human elements (e.g., bullet journaling~\cite{ayobi2018flexible}). We are not interested in purely quantitative data, such as economic indicators or numerical measurements, as they might work better with standard charts and graphs~\cite{kim2019dataselfie}. Instead, we aim to use a questionnaire to collect individual responses. Each data point is personal, and the collection of data points can represent the community.

\subsubsection{Creating humanistic representations of data}

To elicit an emotional response from users, we intend to create humanistic data visualizations. Instead of using traditional visual marks, we resort to using metaphors and rhetoric in our data visualizations. Such visual representations were used in personal data visualizations~\cite{lupi2016dear,kim2019dataselfie} and were also shown to promote better engagement and memorability of the data~\cite{borkin2015beyond,morais2020showing} without necessarily hindering comprehension~\cite{bateman2010useful}.

\subsubsection{Utilizing immersive embodied interactions}
We take one step further to harness the immersive potential of VR environments. They can provide a more visceral data experience by enabling people to use familiar senses, such as sight, sound, and touch, in a real-world-like setting~\cite{donalek2014immersive,romat2020dear,ivanov2018exploration}. Such a heightened sense of immersion can not only enhance the overall enjoyment and engagement but also foster a holistic understanding of the underlying data~\cite{dwyer2018immersive}.

\subsection{System Components: Mapping \& Interaction}

The \toolname{} system consists of three main components.


\subsubsection{Data Collection} 
Qualitative community data is collected using a questionnaire. For the application in this paper, we surveyed students and faculty at a college in the U.S. by asking questions about their backgrounds, personalities (e.g., MBTIs), views of the world (e.g., gloomy and bright), and other opinions and experiences. The responses are collected in a spreadsheet to be fed to the visual mapping module. 

\subsubsection{Visual Mapping}  
Answers to multiple questions in each response are mapped to visual objects and their properties in the virtual environment. For instance, using the college survey, students were visualized as flowers and faculty as trees. Their plastic usage levels are encoded using the number of clouds. Moreover, their MBTI personality type determines the color of one's primary object; the variety of colors in the environment would represent the diversity of personalities in the community. Configuring a new visual mapping can be easily done by supplying new models and specifying mapping parameters in the system.

\subsubsection{Interaction \& Exploration}


Users can freely walk or fly around the environment using a VR device, and switch viewing modes from ground-level to aerial bird's-eye to top-down views. They can open up a visual legend to better understand the visual mapping. In addition, they can perform interactive groupings to facilitate interactive exploration, e.g., organic organization of objects into an intuitive axis-aligned layout of the objects. Finally, a tooltip appears upon touching an object to explain the underlying data. Through this interactive support, \toolname{} not only gains the benefit of immersive visuals but also retains comprehension of the data.






\section{Preliminary User Study}
We conducted a user study with five participants to evaluate \toolname{}. They were asked to perform five tasks using the system, including counting and comparing elements, analyzing relationships, and summarizing trends. After the tasks, they filled out a System Usability Scale survey and engaged in a short interview.

All participants successfully completed the tasks. The overall usability score was 66.5 on a scale from 0 to 100. Overall, participants enjoyed the experience and were engaged throughout. One participant (P2) was particularly engaged by immersion and said, ``It's the feeling of being in a game''. Participants also mentioned that visual aesthetics and physical movements made the experience more positive and meaningful. Some participants also noted that the system could be used in different contexts, such as a social impact application, by humanizing data to encourage empathy and understanding.

However, they often felt overwhelmed by the amount of information presented and had difficulty keeping track of it, which indicates a design trade-off of \toolname{} compared to traditional visualizations. Some participants expressed difficulty in navigation, as well as user controls such as the legend and menu. We observed that people with prior experience in gaming and VR figured out the navigation and interface controls quicker than the others. Based on this feedback, we improved \toolname{}, whose latest version is described in the previous system description.

\section{Conclusion and Future Work}

In this work, we demonstrated a humanistic and immersive approach to data visualization via \toolname{}. For future work, we plan to further improve \toolname{}'s interpretability and scalability through enhanced user interactions (e.g., multi-dimensional filtering) and visual mappings (e.g., adding sound as a variable). We also plan to improve extensibility by making the visual vocabulary fully configurable for different application contexts.
 
\bibliographystyle{ACM-Reference-Format}
\bibliography{references}

\end{document}